# Blended Learning or E-learning?

Maryam Tayebinik
*Faculty of Education, Universiti Teknologi Malaysia, Johor Bahru, Malaysia, Email: ttayebi@gmail.com*

Marlia Puteh
*Language Academy, Universiti Teknology Malaysia, KL Campus, Malaysia, Email: marlia@ic.utm.my*

**Abstract**

ICT or Information and Communication Technology has pervaded the fields of education. In recent years the term "e-learning" has emerged as a result of the integration of ICT in the education fields. Following the application this technology into teaching, some pitfalls have been identified and this have led to the "Blended learning" phenomenon. However, the preference on this new method has been debated quite extensively. The aim of this paper is to investigate the advantages of blended learning over face-to-face instruction through reviews of related literature. The present survey revealed that blended learning is more favorable than pure e-learning and offers many advantages for learners like producing a sense of community or belonging. This study concludes that blended learning can be considered as an efficient approach of distance learning in terms of students' learning experience, student-student interaction as well as student-instructor interaction and is likely to emerge as the predominant education model in the future.

*Keywords*- Blended learning; Online learning; Traditional learning; Educational settings.

## 1. Introduction and Background

"We are living in an ever-changing world" (Sethy, 2008, p.29). Through the past decades, the world of education has been varied by the fast and rapid revolution in computer and the Internet technologies which according to Sethy (2008) "new findings are generated and become established at breathtaking speed" (p.29). This has revolutionized teaching and learning particularly distance education. The arrival of World Wide Web (WWW) has increased the demand for distance education and concepts like online learning or e- learning has emerged, as a result. The system of online learning has been largely used in higher education, and a lot of studies have been done to discover both its strengths and weaknesses (Wang, 2010).

Since e-learning environments present some disadvantages such as inhibiting the socialization process of individuals resulting in lack of face-to-face communication; a new environment has surfaced. This new environment combines the e-learning and the classical learning environments. It has been termed as blended learning, hybrid or mixed learning. The foremost goal of blended instruction was to overwhelm drawbacks of pure online instruction. Since either pure e-learning or traditional learning hold some weaknesses and strengths, it is better to mix the strengths of both learning environments to develop a new method of delivery called blended learning (Azizan, 2010). In view of that, the application of blended instruction has quickly increased because instructors believe that varied delivery methods can increase students' satisfaction from the learning experience as well as their learning outcomes (Lim, & Morris, 2009).

The following section provides extensive reviews of relevant literature about online and blended learning. Different interpretations for the term "Blended Learning" and its usefulness and effectiveness are further discussed.

## 2. What is blended?

The body of literature on blended learning proves that there is no unity on the definition of blended learning. Driscoll (2002) defined blended learning as a combination of instructional methods. On the contrary, Delialioglu and Yildirim (2007) claimed that systematic and strategic combination of ICT tools into academic courses introduces a new way to approach instructional goals. This instructional method has been given many names: blended learning, mediated learning, hybrid instruction, web-assisted instruction or web-enhanced instruction. Delialioglu and Yildirim (2007) and Gülbahar and Madran (2009) believed that blended learning is the same as hybrid instruction, which combines the potentials of web-based training with those of classroom techniques. Likewise, through their study on the transformational potential of blended learning, Garrison and Kanuka (2004) found that blended learning environments seize the values of traditional classes, which improve the effectiveness of meaningful learning experiences. In a more conservative side, Bonk (2004, p.5) cited the three most common definitions of blended learning:

1. A combination of instructional modalities (or delivery media)

2. A combination of instructional methods

3. A combination of online and face-to-face instruction

However, the third definition is mostly accepted by scholars. Picciano (2006), for instance, declared that there are two significant elements in defining blended learning and those are online and face-to-face instructions. Moreover, Rovai and Jordan (2004) claimed that blended learning is a mixture of online learning and classroom that contain some of the facilities of online courses with the presence of face-to-face communication. Other researchers believed that the systems called blended learning integrates face-to-face instruction with computer mediated one (Graham, 2006; Stubbs, Martin & Endlar, 2006; Akkoyunlu & Soylu, 2006).

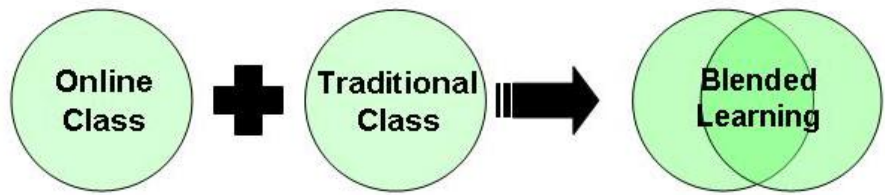

*Figure 1. Blended Learning Environment*

This study applies Colin and Moonen's (2001) definition of blended learning i.e. "a hybrid of traditional face-to- face and online learning so that instruction occurs both in the classroom and online, and where the online component becomes a natural extension of traditional classroom learning" (Colis & Moonen, 2001 cited in Rovai & Jordan, 2004, p.3 ).

3. **Advantages of blended learning**

Parallel with the growing use of ICT in the educational setting, blending learning approach can be contributing tools to complete face to face experiences (Ginns & Ellis, 2009). Besides, blended instruction offers an active learning environment with flexibility in using resources for the students and provides more time for faculty members to spend with learners in small groups or even individually (Oh & Park, 2009). In addition, blended learning has the potential to change students' experiences and outcomes through learning (Davis & Fill, 2007).

Hameed, Badii, and Cullen (2008) in their study considered the efficiency of e-learning when mixed with traditional learning; they concluded that blended learning approach provides the most flexible method to e-learning.

Another advantage of blended learning environments is its potential to offer many sources for learners. Azizan (2010) concluded that utilization of technology in physical classrooms offer extra resources for the students and this is expected to enhance learners' confidence and competence as well as improve the quality of learning. Chen and Jones (2007) outlined other advantages of blended learning such as deep understanding of topics by using web-based resources as well as active participation of students in class. Furthermore, online learning engagement provides an interactive setting for communication among teachers and students in the classroom and may facilitate cooperative activities even beyond the classrooms (Yuen, 2010).

The above discussion has identified the major benefit of applying blended instruction i.e. to overcome the shortcomings of online instruction and exploit various instructional process and delivery strategies in order to increase learners' satisfaction as well as boosting the learning outcomes.

### 4. Blended learning in use

Harrington (2010) coined the combination of traditional classes with online ones as 'hybrid classroom' and stressed that educators are increasingly engage in hybrid classes as they have become aware of the benefits. Moreover, she emphasized that most EFL/ESL students enroll in hybrid classes too.

Usta, and Özdemir (2007) studied students' opinions about blended learning environment and their findings proved that students have generally positive opinions about blended learning environment. The results of the study also proved that high interaction between students and instructor exist in this type of environment. This result supported the findings of Akkoyunlu and Soylu (2006) which indicated high demands for face-to-face interaction in on line learning.

According to Owston et al. (2006, as cited in Bdawi, 2009) there are three rationales for supporting blended learning: fulfilling the learner's needs and motivating critical thinking skills;

the flexibility of blended learning since the learning occurs online and face-to-face and its cost effectiveness.

Dziuban, Hartman & Moskal (2004) in a three-year study between the face- to-face, fully online, and blended teaching methods found that blended teaching always give better success rates than the other two methods. Dowling, Godfrey & Gyles, (2003, as cited in Vignare et al., 2005) investigated the association between students' outcome and hybrid delivery. The results of their study indicated a positive relationship between students' final scores and improved learning outcomes. Moreover, Gómez and Duart (2011) studied a hybrid postgraduate program in a university in Colombia and concluded that students had a very positive opinion of the subjects and the educational model in the program. Similarly, Tselios, Daskalakis, and Papadopoulou, (2011) investigated Greek students' views toward blended learning. The findings obtained showed that both perceived usefulness and simplicity of use have a positive impact on attitude toward using blended learning in the university. Regarding using digital communication tools, Dzakiria, Mustafa and Abu Bakar (2006) claimed that the interaction between students and lecturers as well as scholarly discussion both in synchronous or asynchronous video conference is the privilege offered by blended learning application.

**5. Sense of community in blended learning**

Teachers and students communicate virtually via e-learning and this is the predominant feature of such a learning process, different from traditional classes in which instructors and learners engage in face-to-face interaction (Tayebinik, 2009). Generally, all the terms which describe distance education via computer technology have a unique significance that is learning takes place while teacher and learner are separated. It is assumed that engagement in e-learning and virtual classes hinders e-learners from community interaction. By adding the human interaction to online learning, the educators have considered the human need for socialization which in turn will help the process of learning (Sethy, 2008).

This sense of belonging to a community which its absence in online learning may disturb improvement of common emotions and feelings among learners and instructors. Mc- Millan and Chavis (1986, p.9) offered the following definition of sense of community: "a feeling that members have of belonging, a feeling that members matter to one another and to the group, and a shared faith that members' needs will be met through their commitment to be together".

Rovai and Jordan (2004) studied a causal-comparative design to investigate the relationship of sense of community between fully online, traditional classrooms, and blended higher education learning environments. They found that blended courses create a stronger sense of community among learners than either traditional or fully online courses. Evidently, online or web-based learning environment offers the effectiveness and the flexibility that cannot be guaranteed in a classroom environment while face-to-face classes provide the social communication that students need for learning. So, the integration of these two environments into blended format reserves the advantages of both learning platforms (Akkoyunlu & Soylu, 2006). Hence, it can be concluded that one of the principal benefits of blended learning is providing a sense of community amongst learners (Garrison & Kanuka, 2004).

### 6. Mere online or blended learning?

Proponents of sole e-learning instruction like Lu and Chiou (2010) believe on some benefits of such educational environments like immediate communication, processing learning based on each individual pace, using web technology facilitators (email, chat, video conferencing), etc. Studying through online mode, however, revealed that the feeling of isolation is real and that this negative element is removed through blended learning. The blended learning environment motivates students to participate in online classes more eagerly as they have the opportunity to meet and discuss virtually with their classmates.

Perera (2010) concluded that compared to the virtual learning environment, blended learning offers a more successful learning experience since it contains some aspects of traditional classes. Moreover, mere virtual learning still consists of many problems in the area of education. Based on Hameed, Badii and Cullen (2008), sole e-learning courses is more demanding for instructors and more time commitment is expected of the teacher. Lack of interaction, according to Tinto (1975 as cited in Rovai & Jordan, 2004) will cause frustration and a sense of isolation which leads students to drop out. Tinto (1975 as cited in Rovai & Jordan, 2004) also argued that drop outs are due to inadequate interactions of higher education students with peers and instructors. So, mere online instruction has been denied by many researchers and blended environment has been suggested because of its comprehensible advantages to the educational institutions. Lim, Morris, and Kupritz (2006) concluded that instructions in a blended learning environment seem to be more transparent than using only online delivery format.

Delialioglu and Yildirim (2007) claimed that there are many problems for purely online instruction like limited hardware, software, time, money as well as pedagogical problems. This has lead to a new idea of mixing the benefits of face-to-face courses with the benefits of online courses, known as blended learning. They believed that instructors can support their courses by online exercises, instant online feedback, and creating more valuable learning environments through hypermedia and multimedia.

### 7. Face to face interaction still matters

Face-to-face interaction communicates a lot of facial expressions, body language, tone of voice, and eye contact. Based on Lewis (2006, as cited in Kathy, 2006) facial expressions, body language, and tone of voice are innate. In this regards, the brain needs and expects these more significant channels of information. If these are not available, the brain suffers to communicate and there is a high possibility that a misinterpretation might take place. Lewis (2006) also claimed that if we think we can know someone and embrace this experience through text, we are deceiving ourselves. Visual information and subtle emotions such as winkles and smiles are crucial to communicate anything remotely and these do not exist in online learning. He added that there are many factors that affect human communication that cannot be explained through electronic communication and are more influential than we realize.

Akkoyunlu and Soylu (2006) examined students' view on blended learning environment and discovered that students enjoyed participating in a blended learning environment through which face-to-face classes supplemented with online classes. Moreover, they emphasized on the significance of communication and interaction for successful learning in online education. In another perspective, Rovai (2004), one of the theorists of BL emphasized that designing courses in blended learning is a flexible approach. It provides some conveniences of fully online courses without leaving the face-to-face contact. It can be concluded that the benefits of face-to-face interaction is undeniable and its presence can promote the quality of pure online or traditional classes.

## 8. Conclusion

In a blended learning course i.e. the combination of face-to-face instruction with online platforms, students and teachers engage in using technology for active learning. Furthermore, they are allowed to share their experiences through such a brand environment (Figure 1). In addition, "blended learning provides more productive engagement among students in the online environment and in course content as well" (Ziegler, Paulus, & Woodside, 2006, as cited in Bdawi, 2009, p.6).

Therefore, designing a blended learning environment to reach a harmonious learning equilibrium between face-to- face interaction and online access is essential (Osguthorpe & Graham, 2003, as cited in Bdawi, 2009). Keeping this view in mind, blended environment provides an encouraging situation for both the traditional classrooms and the online settings. In other words, it is a range of delivery methods to meet the course objectives. To sum up, below are the advantages of blended learning over online learning environments as below:

• Increased communication

• Engagement of face-to-face communication

• Sense of community

• Improved academic performance

• Collaborative tasks

• Adequate feedback

• Active participation

• Providing help

• Fun and practical manner of teaching and learning

• Etc…

The above discussions have reviewed the many advantages of blended learning over traditional face-to-face classroom and online instruction. It can be concluded that those activities which are involved in the blended instruction can foster the sense of community belonging and remove the frustration created by mere online environment. The face-to-face element should not be replaced because of the significant effect of body language, tone of voice, facial expressions and eye contact on communication. These two educational settings namely e-learning and face-to-face learning can complement one another for pedagogical application.

This paper has presented the successful application of blended learning in distance learning especially in terms of students' learning experience, student-student interaction and student-instructor interaction. The blended learning approach is likely to emerge as the predominant instructional model in the future.